\documentclass[twocolumn,showpacs,preprintnumbers,amsmath,amssymb]{revtex4}
\UseRawInputEncoding
\usepackage{amssymb}

\usepackage{graphicx}
\usepackage{amsmath}
\usepackage{float}
\usepackage{color}
\usepackage{subeqnarray}
\usepackage{cases}

\begin{document}

\title{{Laser Driven Fluorescence Emission in a Nitrogen Gas Jet at 100 MHz Repetition Rate}
%{Coulomb potential effect in the ultrafast ionization dynamics of noble gas atoms subject to strong elliptically polarized laser fields}

%Strong-field atomic ionization in midinfrared elliptically polarized laser fields

%Quantum interference in correlated electron dynamics from atomic nonsequential double ionization
\footnotetext{$^{*}$ hualq@wipm.ac.cn}
\footnotetext{$^{\dag}$ xjliu@wipm.ac.cn}}
%\footnotetext{$^{\dag}$ xjliu@wipm.ac.cn}}
%\footnotetext{$^{*}$charlywing@wipm.ac.cn \\$^{\dagger}$xjliu@wipm.ac.cn \\$^{\ddagger}$chen$\_$jing@iapcm.ac.cn}}
%\footnotetext{$^{*}$charlywing@wipm.ac.cn \\$^{{\dagger}}$jdlu@wust.edu.cn   }}
%\footnotetext{$^{*}$charlywing@wipm.ac.cn \\$^{\dagger} l \_ j316@163.com$   }}

%\\$^{{\ddagger}}$xjliu@wipm.ac.cn

\author{ Jin Zhang$^{1,2}$, LinQiang Hua$^{1,2,*}$, ShaoGang Yu$^{1,2}$, YanLan Wang$^{1,2}$, MuFeng Zhu$^{1,2}$, ZhengRong Xiao$^{1,2}$, Cheng Gong$^{1,2}$, and XiaoJun Liu$^{1,2,\dag}$ }

\affiliation{$^{1}$ State Key Laboratory of Magnetic Resonance and
Atomic and Molecular Physics, Innovation Academy for Precision Measurement Science
and Technology, Chinese Academy of Sciences, Wuhan 430071, China\\
$^{2}$ University of Chinese Academy of Sciences, Beijing 100049, China}

\date{\today}

%\affiliation{Institute of Applied Physics and Computational
%Mathematics, P. O. Box 8009, Beijing 100088, China}
%\footnote[*] {xjliu@wipm.ac.cn}
%further
% and surviving
%It is shown experimentally that AER is lower at a higher DC electric field, or for an atomic system with higher ionization potential. In contrast, the AER is insensitive to the laser intensity and pulse duration for these high-lying Rydberg states.
\begin{abstract}
We report the fluorescence emission which is driven by femtosecond laser pulses with a repetition rate of 100 MHz and a center wavelength of 1040 nm in a nitrogen gas jet. The experiment is performed in a femtosecond enhancement cavity coupled with high repetition rate laser for the first time to the best of our knowledge. In contrast to previous observation at low repetition rate with a nitrogen gas jet, where the 391 nm radiation was observed but the 337 nm emission was missing, the 337 nm emission is 3 times stronger than the 391 nm emission in our experiment. By examining the dependence of the radiation intensity on the flow rate of the nitrogen gas and the polarization of the pump pulse, the formation mechanism of the N$_2(C^3\Pi_u)$ triplet excited state, i.e., the upper state of the 337 nm emission, is investigated. We attribute the main excitation process to the inelastic collision excitation process, and exclude the possibility of the dissociative recombination as the dominate pathway. The role of the steady state plasma that is generated under our experimental conditions is also discussed.

\end{abstract}

\affiliation{}

%\pacs{33.80.Rv, 33.80.Wz, 42.50.Hz}
%\pacs{****}
\maketitle

\section{INTRODUCTION}
Recent advances in the study on the interaction of ultrashort intense laser pulses with gaseous media have revealed that atmospheric molecules such as N$_{2}$ and O$_2$ can act as nonlinear gain media in air lasing \cite{Dogariu2011,Luo2003,Yao2011,Kartashov2012,Kartashov2013,Mitryukovskiy2014,Yao2014,Ding2014,Mitryukovskiy2015,Li2020,Yao2020,Xie2020}. These air lasers are found to have a variety of promising applications such as remote sensing and standoff spectroscopy \cite{Kasparian2003,Xu2011,Hemmer2011,Yuan2011}. Thus, how the air lasing is generated, and how to increase its intensity, have attracted the attention of many researchers. For N$_2$, the emission at 337 nm, which is due to the transition between the third and the second excited triplet states of neutral nitrogen molecules (2nd positive system of N$_2$), i.e., $C^3\Pi_u \to B^3\Pi_g$, and the emission at 391 nm, which is due to the transition between the second excited state and the ground state of the nitrogen molecular ions (1st negative system of N$_2^+$), i.e., $B^2\Sigma_u^+ \to X^2\Sigma_g^+$, are the two prototype emissions of interest \cite{Talebpour1999,Martin2002}. These two emissions are both assigned to $v=0 \to v'=0$ transitions. They are generally observed with different pump wavelength, different pump polarization state, as well as with seed or not \cite{Kartashov2012,Kartashov2013,Mitryukovskiy2014,Yao2014,Ding2014,Mitryukovskiy2015}.

Because the direct photonic excitation of the triplet state N$_2(C^3\Pi_u)$ is a spin forbidden process, three different mechanisms have been proposed to be responsible for the population of the $C^3\Pi_u$ state, i.e., the upper state of the 337 nm  emission. However, a conclusive agreement has not yet been reached. The first scheme was proposed by Xu \emph{et al.} at 2009 \cite{Xu2009}. They suggested that the population of the $C^3\Pi_u$ state is a dissociative recombination process through the following two steps: $N_2^+ + N_2 \to N_4^+$ followed by $N_4^+ + e^- \to N_2(C^3\Pi_u) + N_2$. The second scheme proposed an intersystem crossing (ISC) scenario from its singlet excited states, i.e., $N_2 + hv \to N_2^\ast$(singlet excited states), $N_2^\ast + M \to N_2(C^3\Pi_u) + M$ \cite{Arnold2012,Li2016,Li2017,Talebpour2001}. The third scheme is a direct inelastic collision excitation process by electrons: $N_2(X^1\Sigma_g) + e^- \to N_2(C^3\Pi_u) + e^-$ \cite{Mitryukovskiy2014,Yao2014,Ding2014,Mitryukovskiy2015}. It was suggested that a large number of electrons with kinetic energies over 14 eV can be produced with a circularly polarized laser field. Such an energy is sufficient for the excitation of nitrogen molecules from their ground states to the excited triplet states through inelastic collision excitations \cite{Fons1996}.

Note that in the aforementioned three mechanisms, the production of the photoelectrons were given a pivotal role in the population of the $C^3\Pi_u$ states. Therefore, studying the emission of 337 nm under various electronic environments, such as different electron density, different kinetic energy (temperature), etc., provides valuable clues to identify the role of each mechanism. On the other hand, control over the density and temperature of the electrons are beneficial to increase the efficiency of the population of the $C^3\Pi_u$ states and thus the emission intensity of the air lasing \cite{Sprangle2011,Itikawa2006}.

In the past two decades, femtosecond enhancement cavity (fsEC) coupled with high repetition rate lasers (typically $\ge$10 MHz) not only helps to realize the frequency comb in the extreme ultraviolet (XUV) region \cite{Jones2005,Gohle2005,Lee2011,Mills2012,Ozawa2015,Zhang2020}, but also provides a new platform to study strong field atomic and molecular dynamics with an unprecedented high repetition rate, and accordingly, with a high signal-to-noise ratio and statistics \cite{Yost2009,Benko2015,Hogner2017,Zhang2020,Corder2018,Na2019}. Inside the fsEC which is coupled with high repetition rate laser, the plasma generated in the focal volume by the first pulse does not have time to be cleared before the successive laser pulses arrive and generate even more plasma \cite{Allison2011,Porat2018}. Consequently, a higher density of steady-state plasma is formed at the focus of the fsEC compared to the conventional experiment with 1 kHz laser under the same laser intensity, thus providing a novel electronic environment to the interaction dynamics between laser and matter in the focus region \cite{Allison2011,Porat2018}.

In this paper, we employ the fsEC to study the fluorescence emission in a nitrogen gas jet which is driven by femtosecond laser pulses with a repetition rate of 100 MHz and a center wavelength of 1040 nm. To the best of our knowledge, this is the first experiment to study the fluorescence dynamics with a repetition rate of 100 MHz. At such high repetition rate, the high density of electrons that is generated by the steady state plasma leads to a high collision rate between electrons and nitrogen molecules and other particles. Thus, an increase of the 337 nm emission due to a stronger collision is expected. In addition, we discuss the formation mechanism of the $C^3\Pi_u$ excited triplet state by examining the dependence of the intensity of 337 nm emission on the flow rate of the nitrogen gas and the polarization of the pump laser. Our results reveal that the generated steady state plasma plays an important role in the population of the $C^3\Pi_u$ states.

\section{EXPERIMENTAL SETUP}

In our experiment, an Yb-doped fiber laser system (Active Fiber systems GmbH) is utilized as the driving laser. It delivers pulses with a repetition rate of 100 MHz, a central wavelength of 1040 nm, and a maximum output pulse energy of 1 $\mu$J. This pulse energy is rather low so that it is very hard to generate plasma in nitrogen molecules under normal focal configuration. Thus, we use a passive optical cavity to enhance the pulse energy of the pump laser, as shown in Fig. 1. The details of our femtosecond enhancement cavity has been described in Ref. \cite{Zhang2019,Zhang2020}. Here, only a brief summary is given. The fsEC is a six-mirror bowtie cavity, which is composed of an input coupler (IC) with R$\approx$99.2\%, and five high reflection mirrors with R$\approx$99.95\%. With this design, the finesse of the cavity is expected to be about 600 and the theoretical buildup is 290. By the aid of the Pound-Drever-Hall (PDH) technique, the cavity length is locked to maintain the maximum intracavity power. A quarter-wave plate is inserted before the input coupler of the enhancement cavity to change the laser polarization between linear and circular. With a pump power of 34 W, we have achieved an intra-cavity power of 6.38 kW and the corresponding buildup is 187. The pulse duration is measured to be $\sim$250 fs by a frequency-resolved optical gating (FROG). The focus radius is estimated to be 8 $\mu m$ (vertical) $\times$ 16 $\mu m$ (horizontal) from an ABCD matrix analysis of the cavity. The peak intensity in the focus region is evaluated to be $6.4\times10^{13}$ W/cm$^2$. The ponderomotive potential $U_p$ of the electron in a linearly polarized laser field is about 6.4 eV. We inject nitrogen gas at the intracavity focus using a glass nozzle with an aperture of 150 $\mu m$ and a backing pressure of 2 bar. The background pressure for the vacuum chamber is $\sim$0.74 Pa. Because of the tight focus of the cavity, the forward emission is collimated by an f = 750 mm CaF$_2$ lens 0.5 m after MA. After passing through a color glass filter (KG3) which blocks the strong 1040 nm signal, the forward emission is focused by an f = 100 mm fused silica lens and measured by a spectrometer (Ocean Optics maya 2000 pro). The integration time of the spectrometer is $\sim$500 ms.

\begin{figure}[htbp]
\centering
\includegraphics[width=3.2in]{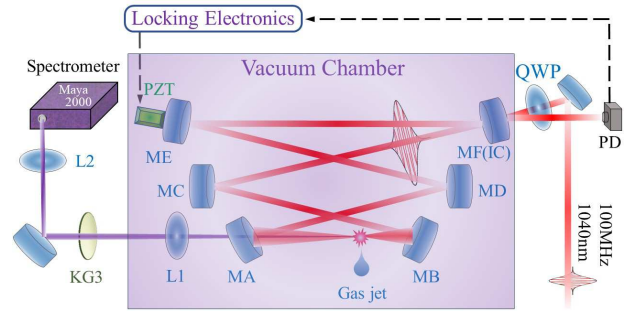}
\vspace{0.1in}\caption{Schematic view of our experimental setup. QWP: quarter-wave plate, PZT: piezo transducer, IC: input coupler, MA-MF: mirror A-mirror F, L1: f = 750 mm CaF$_2$ lens, L2: f = 100 mm fused silica lens, PD: photodiode.}
\end{figure}

\section{RESULTS}
Fig. 2 shows a typical forward emission spectrum under linear polarization pump laser in our experiment. The emission peaks centered at 315.8 nm, 337.0 nm, 380.3 nm and so on have been well identified as the transitions between the triplet states $C^3\Pi_u$ and $B^3\Pi_g$ of neutral nitrogen molecules, while the 391.3 nm line (named as 391 nm line in this paper) is identified as the transitions between the second excited state $B^2\Sigma_u^+$ and ground state $X^2\Sigma_g^+$ of nitrogen molecular ions \cite{Talebpour1999,Martin2002}. The different initial and final vibrational quantum numbers are denoted in Fig. 2. A saturated peak around 346 nm is the 3rd harmonic of the fundamental laser. The emission intensity at 337 nm is about 3 times stronger than the 391 nm emission. The bandwidth of the 337 nm emission was measured to be 0.2 nm, which corresponds to the spectral resolution of our spectrometer. This bandwidth is very close to the theoretical value of the transition between the $C^3\Pi_u$ and $B^3\Pi_g$ state, which was determined to be $\sim$0.1 nm \cite{Hariri2014}. It is worth of noting that, our observed spectra are very different from the ones obtained by Plenge \emph{et al.}, where no $C^3\Pi_u$ state emission was observed with a gas jet \cite{Plenge2009}.

\begin{figure}[htbp!]
\centering\includegraphics[width=3.2in]{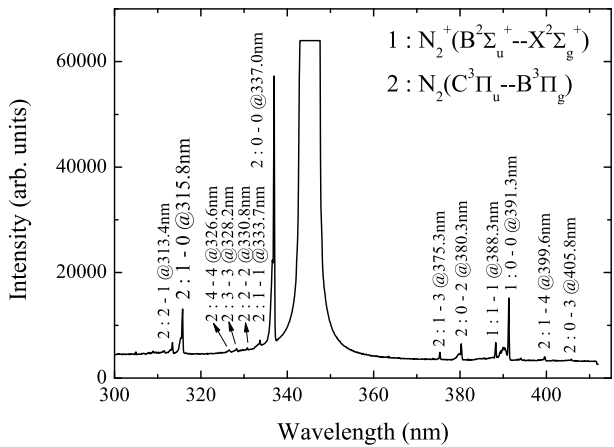}
\vspace{0.1in}\caption{Spectra of the forward-propagating UV optical signal. In the (v-v') transitions, v and v' denote the vibrational levels of upper and lower electronic states, respectively.}
\end{figure}

We then measured the polarization of the forward emission at 337 nm under linear polarization pump laser. Before the measurement, an achromatic half wave plate (HWP) and a Glan-Taylor prism are inserted before the spectrometer, and they are not shown in Fig. 1. In the experiment, we record the intensity of the transmitted 337 nm radiation as a function of the rotation angle of the HWP. The result is displayed in Fig. 3. No obvious intensity dependence is observed, indicating that the emission at 337 nm is not polarized.

\begin{figure}[htbp!]
\centering\includegraphics[width=3.2in]{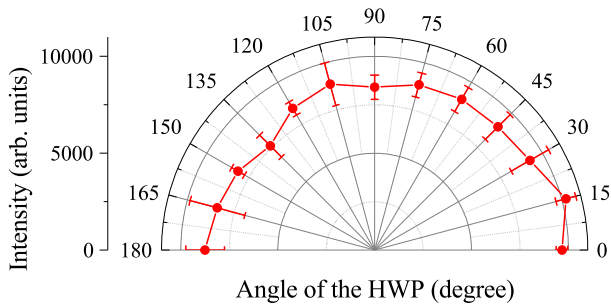}
\vspace{0.1in}\caption{The intensity of the forward emission signal at 337 nm as a function of the rotation angle of the HWP.}
\end{figure}

We also measured the intensity dependence of the 337 nm forward emission on the gas flow rate, as shown in Fig. 4. The result shows that with an increase of the gas flow rate, the forward emission signals at 337 nm increase with a linear trend.

\begin{figure}[htbp!]
\centering\includegraphics[width=3.2in]{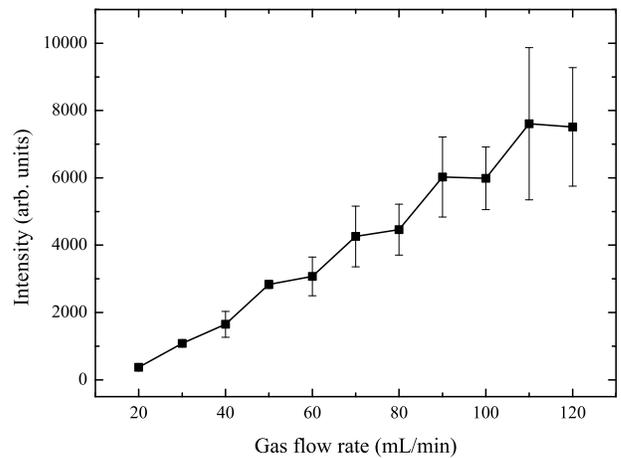}
\vspace{0.1in}\caption{The forward emission intensity at 337 nm as a function of the gas flow rate.}
\end{figure}

Finally, we test the influence of the polarization of the pump laser on the intensity of the 337 nm radiation. The 337 nm emission is measured for different polarization of the incident laser with the same intra-cavity power, as shown in Fig. 5. Our result shows that linearly polarized pulses generate a radiation at 337 nm with an intensity of about 3 times stronger than that obtained with circular polarization.

\begin{figure}[hbp!]
\centering\includegraphics[width=3.2in]{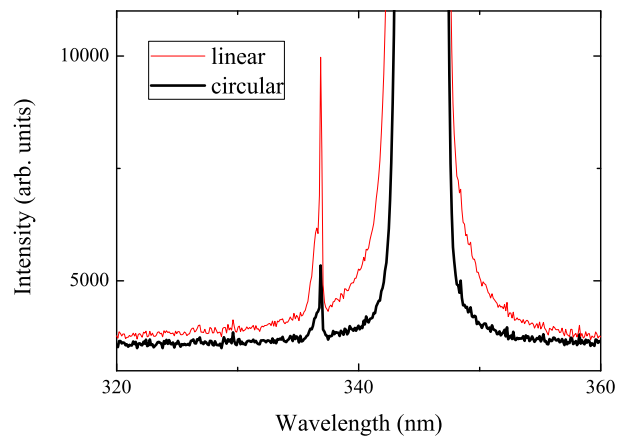}
\vspace{0.1in}\caption{Forward emission spectra when pumped with linearly (red thin) and circularly (black thick) polarized pulses.}
\end{figure}

\section{DISCUSSIONS}
Fig. 2 shows that the emission from the $C^3\Pi_u$ state has been clearly observed, which is very different from the results obtained by Plenge \emph{et al.}, where no $C^3\Pi_u$ state emission was observed with a gas jet \cite{Plenge2009}. In their experimental work performed with 1 kHz laser, it was concluded that the disappeared fluorescence was due to the collision free environment in a gas jet, leading to no population of the $C^3\Pi_u$ state \cite{Plenge2009}. However, under our experimental condition of high repetition rate (i.e., 100 MHz), the collision between electrons and nitrogen molecules can't be neglected even though a gas jet is adopted. In our case, the molecular density at the focus is $\sim 10^{18}$ cm$^{-3}$\cite{Mills2012}. The speed is calculated to be about $2\times10^{6}$ m/s for the electrons with an energy of 12.8 eV (corresponding to 2$U_p$ for our experimental laser intensity). Thus, the mean collision time between electrons and nitrogen molecules can be estimated to be around 4 ps in our gas jet. And note that, the peak intensity in our experiment is much lower than the experiment of Ref. \cite{Plenge2009}. In their experiment, the peak intensity is about $3\times10^{14}$ W/cm$^2$, thus it is reasonable to consider that nearly all of the nitrogen molecules are ionized \cite{Guo1998}, resulting in a very low density of neutral molecules. Ultimately, no $C^3\Pi_u$ state emission can be observed in their experiment.

Furthermore, in our experiment with a high-repetition rate laser, the collision probability between electrons and N$_2$ molecules becomes higher compared with 1 kHz system. The main difference induced by the high repetition rate laser is that the plasma formed by the previous laser pulse doesn't have enough time to decay before the next pulse arrives, and thus a steady state plasma is formed \cite{Allison2011,Porat2018}. To gain a deeper understanding on this effect, an estimation is used to qualitatively understand the formation of the steady state plasma. Suppose the speed of the nitrogen molecular beam is v = 658 m/s \cite{Miller1988} and the time interval between two successive laser pulses is t = 10 ns, the plasma generated by the previous pulse moves only s = 6.58 $\mu m$ when the next pulse arrives and thus, can not dissipate out of the laser focus region (with a diameter of d = 16 $\mu m$ in our fsEC according to the design). Consequently, the intra-cavity pulse will interact with the gas medium several times and generate a higher density of steady-state plasma, which would enhance the collision between nitrogen molecules and electrons. Two different contributions to the plasma density are taken into consideration: $\eta_{pulse}$ created by a single laser pulse and $\eta_{steady}$ which persists from pulse to pulse. To unravel their relative contributions, we have developed a numerical model to decouple their respective role. The empirical formula of Tong and Lin \cite{Tong2005} is used to construct the cycle-averaged and peak ionization rates $\bar{w}(\epsilon)$ and $w(\epsilon)$, respectively. The ionization fraction during the pulse, $\eta_{pulse}$, is then calculated via
\begin{equation}
\eta_{pulse}(t)=1-exp[-\int_{-\infty}^{t}dt\bar{w}(\epsilon)],
\end{equation}
where $\epsilon$ is the electric field envelope. The steady-state ionization $\eta_{steady}$ is calculated from a balance between the plasma creation and the decay per round trip:
\begin{equation}
\eta_{steady}|_{n+1}=\alpha*\eta_{pulse}+(1-\alpha)*[\eta+(1-\eta)*\eta_{pulse}],
\end{equation}
where $\eta_{steady}|_{n+1}$ is the fraction of the steady state plasma after the n+1 pulse, $\alpha$ = s/d is the new gas ratio, $\eta=-k_p*\eta_{steady}|_n$, is the decay, and $k_p$ is a constant. In Eq.(2), the first term $\alpha*\eta_{pulse}$ is the plasma fraction of the new gas filled in the focus region ionized by the n+1 pulse. The second term $(1-\alpha)*\eta$ is the residual plasma fraction after decay. And the third term $(1-\alpha)*(1-\eta)*\eta_{pulse}$ is the plasma fraction ionized by the n+1 pulse of the old gas, i.e., the gas which has not dissipated out of the focus region before the n+1 pulse arrives. Using the parameters described in the previous section, the $\eta_{pulse}$ and $\eta_{steady}$ are calculated to be 0.17 and 0.30, respectively. The steady-state plasma density in a high-repetition rate system is about twice of the 1 kHz system. The higher plasma density further suggests that the collision cannot be eliminated under our experimental condition even though a gas jet is adopted.

Besides the strong emission of the 337 nm line, we also noticed the disappearance of the 357 nm line, corresponding to the transition between the $C^3\Pi_u$ (v' = 0) and $B^3\Pi_g$ (v = 1) states, as shown in Fig. 2. This observation seems abnormal since this transition shares the same upper state, i.e., $C^3\Pi_u$ (v' = 0) state, with the 337 nm transition. Comparing our results with other experiments under similar condition suggests that the 357 nm emission is wavelength dependent, i.e., the 357 nm emission disappears with $\sim$1 $\mu m$ pump \cite{Kartashov2013}, while it appears when 800-nm laser is used as a pump \cite{Mitryukovskiy2014,Yao2014,Ding2014,Mitryukovskiy2015}. It is probable that the vibrational populations of the $C^3\Pi_u$ and $B^3\Pi_g$ states of the nitrogen molecules are wavelength dependent and further investigation on the vibrational resolved dynamics is expected.

%Similar results have also been reported by Kartashov et. al \cite{Kartashov2013}. The similarity of these two experiments is that the wavelengths of the pump laser are both around 1 $\mu m$, so that the wavelength of the third harmonic of the pump laser locates in the range between 337 to 357 nm. In contrast, in other experiments which driven by 800-nm laser, the 357 nm line can be observed although its third harmonic is 266 nm, far away from 337 nm \cite{Mitryukovskiy2014,Yao2014,Ding2014,Mitryukovskiy2015}. This indicates that the vibrational populations of the $C^3\Pi_u$ state of the nitrogen molecules is wavelength dependent and more investigate on the vibrational resolved dynamics is expected.

Fig. 3 shows that the emission at 337 nm is not polarized. This is not surprising since the polarization of the unseeded Amplified Spontaneous Emission (ASE) depends strongly on the experimental condition. In an experiment which employed a strong 1053 nm picosecond laser as the pump \cite{Kartashov2013}, the 337-nm emission was observed to be linearly polarized (parallel to the polarization direction of the pump light), suggesting that the emission was seeded by the short-wavelength spectral tail of the third harmonic of the pump. It is noteworthy that the gain built-up time is $\sim$4 ps for 1 bar nitrogen gas \cite{Yao2014,Ding2014}. Thus, the third harmonic produced by a 10-ps pump laser is still present when the population inversion becomes positive and it is available to the seed of the 337 nm emission \cite{Kartashov2013}. In our experiment, because the pulse duration of the pump laser is hundreds of femtoseconds, which is much shorter than the gain built-up time. Thus, the third harmonic isn't available to seed the 337-nm emission. Our observation is similar to the case in Ref. \cite{Ding2014}, where the unseeded ASE is also not polarized.

%Moreover, it should be pointed out that the seed can't be amplified when the central wavelength of the seed is tuned away from 337 nm \cite{Yao2014}. But in Kartashov's experiment, the third harmonic of the pump, which is centered at 351 nm, seeds the 337-nm emission \cite{Kartashov2013}. Thus, it is likely that the 337 nm seed are generated in the coherent vibrational Raman scattering process \cite{Xu2018}, since the 351 nm is as strong as 700$\mu$J in their experiment.The generated 337-nm Raman seed signal is then amplified in the population-inverted nitrogen molecules, resulted in a polarized 337 nm lasing.

Fig. 4 shows that the intensity of the 337 nm emission depends linearly on the gas flow rate. Suppose the dissociative recombination mechanism is the dominant pathway, i.e., $N_2^+ + N_2 \to N_4^+$ followed by  $N_4^+ + e^- \to N_2(C_3\Pi_u) + N_2$, the 337 nm emission should increase at least quadratically with the gas flow rate, since N$_2^+$, N$_2$, and $e^- $ all increase linearly with gas flow rate. Thus, we eliminate the possibility that the dissociative recombination mechanism as the dominant pathway under our experimental condition. On the other hand, the intersystem crossing mechanism requires the presence of a high concentration of heavy atoms such as He for collision excitation \cite{Arnold2012}, which is not the case in pure nitrogen sample. For inelastic collision excitation mechanism, i.e., $N_2(X^1\Sigma_g) + e^- \to N_2(C^3\Pi_u) + e^-$, the concentration of N$_2$ in ground state and the ionized electron both depend linearly on pressure. Therefore, in consideration of the quenching of the triplet excited N$_2$, the 337 nm emission may increase linearly with the gas flow rate.

%the intersystem crossing mechanism, the $N_2^*$ in excited singlet state and the collision partner M both depends linearly on pressure. However, in consideration of the quenching of excited $N_2^+$, the 337 nm emission may increase linearly as the gas flow rate. Similar conclusion can also be made for the direct electron collision mechanism.

%The dependence of the 391 nm emission on the gas flow rate depicts a tiny increase, as shown in Fig. 4. Similarly, more $N_2^+$ on the $B^2\Sigma_u^+$ state are generated with increasing the gas flow rate, resulting in a stronger 391 nm emission. On the other hand, free electron are found to quench $N_2^+$ on the $B^2\Sigma_u^+$ state with high efficiency. With time-resolved fluorescence spectroscopy, Lei et al. noticed that the decay of $N_2^+$ on the $B^2\Sigma_u^+$ state consists of two component \cite{Lei2017}. The fast one with a few hundred picoseconds is due to collision with free electron, while the slow one with $\sim$10 ns is ascribe to collision with $N_2$ molecule or spontaneous decay. The fast quenching of the $N_2^+$ on the $B^2\Sigma_u^+$ state is probably the reason why the 391 nm emission shows only a tiny increase with increasing the gas flow rate.

Fig. 5 shows that the emission at 337 nm is stronger with linearly polarized laser than with circularly polarized laser. This observation is in agreement with two previous reports in the low intensity regime \cite{Mitryukovskiy2015,Danylo2019}, where a stronger 337 nm emission is also observed with linearly polarized laser. On the other hand, these results are different from those experiments which are operated with high laser intensity, where circularly polarized laser generates stronger 337 nm emission \cite{Mitryukovskiy2014,Yao2014,Ding2014,Mitryukovskiy2015}.The key difference induced by the polarization of the laser is that, with linearly polarized laser pulses, most free electrons are left with low kinetic energy at the end of the pump pulse because they are alternately accelerated and decelerated by the laser field during every optical cycle. While with a circularly polarized laser field, electrons are always accelerated away from their parent molecular ions and they acquire an average energy of $\sim$2$U_p$ ($U_p$ is the ponderomotive potential) at the end of the laser pulse \cite{Bucksbaum1986,Corkum1989}. Thus, in the high intensity regime (typically $\ge1\times10^{14}$ W/cm$^2$ ), a large number of electrons with kinetic energies over 14 eV can be produced with a circularly polarized laser field and it is more likely to excite the N$_2$ molecule to its $C^3\Pi_u$ state \cite{Mitryukovskiy2014,Yao2014,Ding2014,Mitryukovskiy2015}. This mechanism explains the result in the high intensity regime very well, but leaves the result in the low intensity regime as a puzzle since electrons that are generated with linear polarized laser don't have enough energy to excite N$_2$ to the $C^3\Pi_u$ state. Recently, Liu \emph{et al.} proposed the multiple collision as a possible candidate mechanism \cite{Danylo2019}. However, it is worthy to note that electrons generated due to recollisions have not been taken into account in the above discussions \cite{Mitryukovskiy2014,Yao2014,Ding2014,Mitryukovskiy2015,Danylo2019}. Indeed, the recollision between the electron and its parent ion greatly increases the electron kinetic energy and electron yields in the plateau region. This may lead to an increase of the intensity of the 337 nm emission, especially in the case of a linearly polarized laser field \cite{Zheng2020}. On the other hand, in the case of the circularly polarized laser, the probability of recollision between the electron and its parent ion is low, and the high energy electrons due to rescattering may be neglected. We have theoretically calculated the electron energy distribution in circularly polarized and linearly polarized laser fields in the context of strong field approximation \cite{Becker2002}. Under our laser intensity, the electron yield in the linearly polarized laser field is higher than that in the circularly polarized laser field at around 14 eV, as shown in Fig. 6. That is probably why the 337 nm radiation is stronger with a linearly polarized laser as a pump under our experimental condition.

\begin{figure}[htbp!]
\centering\includegraphics[width=3.2in]{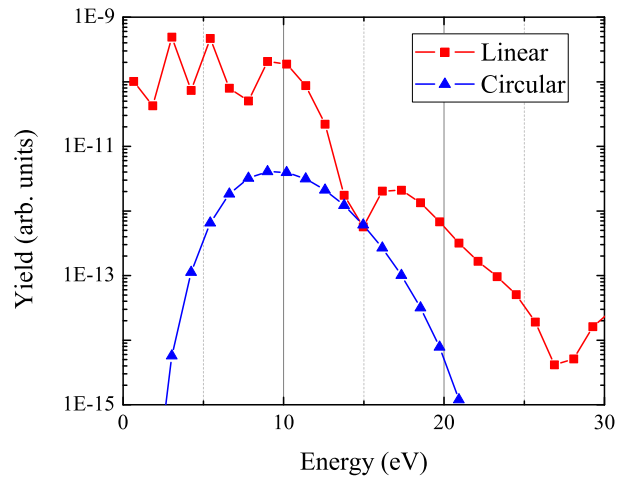}
\vspace{0.1in}\caption{Theoretically predicted electron energy distribution in circularly polarized (blue triangle) and linearly polarized (red square) laser fields with an intensity of $6.4\times10^{13}$ W/cm$^2$}
\end{figure}

Before concluding, we would like to recall the observations among different groups and try to gain a deeper insight into the underlying excitation mechanism of the $C^3\Pi_u$ state molecules. So far, different groups have noticed the fast population (in a few picosecond) of the $C^3\Pi_u$ state molecules \cite{Kartashov2013,Yao2014,Danylo2019,Xu2009,Xu2011APL}. This fast population can only be achieved by collision with electrons \cite{Danylo2019}. Under such circumstances, it is reasonable to eliminate the ISC mechanism and the dissociative recombination mechanism as the predominant candidates. For these mechanisms, at least one step takes much longer than a few picoseconds \cite{Danylo2019,Sprengers2004,Peng2019}. Besides, as also demonstrated in our experiment with polarization dependence and flow rate dependence measurements, these mechanisms are in contradiction to our observations as well. Thus, we would like to attribute the inelastic collision excitation process with electron as the most possible pathway to populate the $C^3\Pi_u$ state and this process is enhanced by the steady state plasma with a higher rate of collision for a high repetition rate laser system.

\section{Conclusion}
In conclusion, we report a fluorescence emission with a record-high 100 MHz repetition rate by the aid of a dedicated fsEC. This fluorescence is driven by femtosecond laser pulses at 1040 nm in a nitrogen gas jet. At such high repetition rate, plasma accumulates at the focus resulting in a higher plasma density, which has the potential to be beneficial to increase the intensity of 337 nm emission. We discuss the formation mechanism of the excited triplet state by examining the dependence of the intensity of the 337 nm emission on the flow rate of the nitrogen gas and the polarization of the pump laser. These results eliminate the possibility that the dissociative-recombination is the dominant pathway, and suggest that inelastic collision excitation mechanism is the major route to populate the $C^3\Pi_u$ state of nitrogen molecules under our experimental condition. Our study shows that a versatile fsEC, in addition to its popular employment in the XUV comb production, can also be used as a unique tool to inspect the role of each mechanism in the formation dynamics of air lasing.

\section*{ACKNOWLEDGMENTS}

This work is supported by the National Natural
Science Foundation of China (Nos. 11674356, 11804374, 12004391 and 11527807), the Strategic Priority Research
Program of the Chinese Academy of Sciences (No. XDB21010400), the Science and Technology Department of Hubei Province (No. 2020CFA029), and the K. C. Wong Education Foundation.


\begin{thebibliography}{0}
\expandafter\ifx\csname natexlab\endcsname\relax\def\natexlab#1{#1}\fi
\expandafter\ifx\csname bibnamefont\endcsname\relax
  \def\bibnamefont#1{#1}\fi
\expandafter\ifx\csname bibfnamefont\endcsname\relax
  \def\bibfnamefont#1{#1}\fi
\expandafter\ifx\csname citenamefont\endcsname\relax
  \def\citenamefont#1{#1}\fi
\expandafter\ifx\csname url\endcsname\relax
  \def\url#1{\texttt{#1}}\fi
\expandafter\ifx\csname urlprefix\endcsname\relax\def\urlprefix{URL }\fi
\providecommand{\bibinfo}[2]{#2}
\providecommand{\eprint}[2][]{\url{#2}}

\end{thebibliography}


\begin{thebibliography}{99}
%1
 \bibitem{Dogariu2011}
 A. Dogariu, J. B. Michael, M. O. Scully, and R. B. Miles,  High-Gain Backward Lasing in Air, Science \textbf{331}, 442 (2011).

 \bibitem{Luo2003}
 Q. Luo, W. Liu, and S. L. Chin,  Lasing action in air induced by ultra-fast laser filamentation,  Appl. Phys. B \textbf{76}, 337 (2003).

  \bibitem{Yao2011}
 J. Yao, B. Zeng, H. Xu, G. Li, W. Chu, J. Ni, H. Zhang, S. L. Chin, Y. Cheng, and Z. Xu, High-brightness switchable multiwavelength remote laser in air, Phys. Rev. A \textbf{84}, 051802(R) (2011)

 \bibitem{Kartashov2012}
 D. Kartashov, S. Alisauskas, G. Andriukaitis, A. Pugzlys, M. Shneider, A. Zheltikov, S. L. Chin, and A. Baltuska, Free-space nitrogen gas laser driven by a femtosecond filament,  Phys. Rev. A \textbf{86}, 033831 (2012).

 \bibitem{Kartashov2013}
 D. Kartashov, S. Alisauskas, A. Baltuska, A. Schmitt-Sody, W. Roach, and P. Polynkin,  Remotely pumped stimulated emission at 337 nm in atmospheric nitrogen,  Phys. Rev. A \textbf{88}, 041805(R) (2013).
%6
 \bibitem{Mitryukovskiy2014}
 S. Mitryukovskiy, Y. Liu, P. Ding, A. Houard, and A. Mysyrowicz,  Backward stimulated radiation from filaments in nitrogen gas and air pumped by circularly polarized 800 nm femtosecond laser pulses, Opt. Express \textbf{22}, 12750 (2014).

 \bibitem{Yao2014}
 J. Yao, H. Xie, B. Zeng, W. Chu, G. Li, J. Ni, H. Zhang, C. Jing, C. Zhang, H. Xu, Y. Cheng, and Z. Xu,  Gain dynamics of a free-space nitrogen laser pumped by circularly polarized femtosecond laser pulses,  Opt. Express \textbf{22}, 19005 (2014).

 \bibitem{Ding2014}
 P. Ding, S. Mitryukovskiy, A. Houard, E. Oliva, A. Couairon, A. Mysyrowicz, and Y. Liu,  Backward Lasing of Air plasma pumped by Circularly polarized femtosecond pulses for the saKe of remote sensing (BLACK), Opt. Express \textbf{22}, 29964 (2014).

 \bibitem{Mitryukovskiy2015}
  S. Mitryukovskiy, Y. Liu, P. Ding, A. Houard, A. Couairon, and A. Mysyrowicz,  Plasma Luminescence from Femtosecond Filaments in Air: Evidence for Impact Excitation with Circularly Polarized Light Pulses, Phys. Rev. Lett. \textbf{114}, 063003 (2015).

 \bibitem{Li2020}
 H. Li, E. L$\ddot{o}$tstedt, H. Li, Y. Zhou, N. Dong, L. Deng, P. Lu, T. Ando, A. Iwasaki, Y. Fu, S. Wang, J. Wu, K. Yamanouchi, and H. Xu, Giant Enhancement of Air Lasing by Complete Population Inversion in N$_2^+$, Phys. Rev. Lett. \textbf{125}, 053201 (2020).

  \bibitem{Yao2020}
 J. Yao, and Y. Cheng,  Air Lasing: Novel Effects in Strong Laser Fields and New Technology in Remote Sensing, Chin. J. Las. \textbf{47}, 0500005 (2020).

  \bibitem{Xie2020}
 H. Xie, H. Lei, G. Li, Q. Zhang, X. Wang, J. Zhao, Z. Chen, J. Yao, Y. Cheng, and Z. Zhao, Role of rotational coherence in femtosecond-pulse-driven nitrogen ion lasing, Phys. Rev. Res. \textbf{2}, 023329 (2020).
%11
 \bibitem{Xu2011}
 H. Xu, and S. L. Chin,  Femtosecond Laser Filamentation for Atmospheric Sensing,  Sensors \textbf{11}, 32 (2011).

 \bibitem{Kasparian2003}
 J. Kasparian, M. Rodriguez, G. Mejean, J. Yu, E. Salmon, H. Wille, R. Bourayou, S. Frey, Y. B. Andre, A. Mysyrowicz, R. Sauerbrey, J. P. Wolf, and L. Woste,  White-light filaments for atmospheric analysis, Science \textbf{301}, 61 (2003).

 \bibitem{Hemmer2011}
 P. R. Hemmer, R. B. Miles, P. Polynkin, T. Siebert, A. V. Sokolov, P. Sprangle, and M. O. Scully,  Standoff spectroscopy via remote generation of a backward-propagating laser beam,  Proc. Natl. Acad. Sci. U. S. A. \textbf{108}, 3130 (2011).

 \bibitem{Yuan2011}
 L. Yuan, A. A. Lanin, P. K. Jha, A. J. Traverso, D. V. Voronine, K. E. Dorfman, A. B. Fedotov, G. R. Welch, A. V. Sokolov, A. M. Zheltikov, and M. O. Scully,  Coherent Raman Umklappscattering, Laser Phys. Lett. \textbf{8}, 736 (2011).

 \bibitem{Talebpour1999}
 A. Talebpour, S. Petit, and S. L. Chin,  Re-focusing during the propagation of a focused femtosecond Ti : Sapphire laser pulse in air,  Opt. Commun. \textbf{171}, 285 (1999).
%16
 \bibitem{Martin2002}
 F. Martin, R. Mawassi, F. Vidal, I. Gallimberti, D. Comtois, H. Pepin, J. C. Kieffer, and H. P. Mercure,  Spectroscopic study of ultrashort pulse laser-breakdown plasmas in air,  Appl. Spectrosc. \textbf{56}, 1444 (2002).

\bibitem{Xu2009}
H. Xu, A. Azarm, J. Bernhardt, Y. Kamali, and S. L. Chin,  The mechanism of nitrogen fluorescence inside a femtosecond laser filament in air,  Chem. Phys. \textbf{360}, 171 (2009).

\bibitem{Arnold2012}
B. R. Arnold, S. D. Roberson, and P. M. Pellegrino,  Excited state dynamics of nitrogen reactive intermediates at the threshold of laser induced filamentation,  Chem. Phys. \textbf{405}, 9 (2012).

 \bibitem{Li2016}
  S. Li, L. Sui, A. Chen, Y. Jiang, D. Liu, Z. Shi, and M. Jin,  Angular distribution of plasma luminescence emission during filamentation in air,  Phys. Plasmas \textbf{23}, 5 (2016).

 \bibitem{Li2017}
  S. Li, Y. Jiang, A. Chen, L. He, D. Liu, and M. Jin,  Revisiting the mechanism of nitrogen fluorescence emission induced by femtosecond filament in air,  Phys. Plasmas \textbf{24}, 5 (2017).
%21
\bibitem{Talebpour2001}
  A. Talebpour, M. Abdel-Fattah, A. D. Bandrauk, and S. L. Chin,  Spectroscopy of the gases interacting with intense femtosecond laser pulses,  Laser Phys. \textbf{11}, 68 (2001).

 %\bibitem{Talebpour19961}
  %A. Talebpour, C. Y. Chien, and S. L. Chin,  Population trapping in rare gases,  J. Phys. B: At. Mol. Opt. Phys. \textbf{29}, 5725 (1996).


 %\bibitem{Talebpour19971}
  %A. Talebpour, C. Y. Chien, Y. Liang, S. Larochelle, and S. L. Chin,  Non-sequential ionization of Xe and Kr in an intense femtosecond Ti:Sapphire laser pulse,  J. Phys. B: At. Mol. Opt. Phys. \textbf{30}, 1721 (1997).

 %\bibitem{Talebpour19962}
 % A. Talebpour, Y. Liang, and S. L. Chin,  Population trapping in the CO molecule,  J. Phys. B: At. Mol. Opt. Phys. \textbf{29}, 3435 (1996).

 %\bibitem{Talebpour19972}
  %A. Talebpour, S. Larochelle, and S. L. Chin,  Non-sequential and sequential double ionization of NO in an intense femtosecond Ti:Sapphire laser pulse,  J. Phys. B: At. Mol. Opt. Phys. \textbf{30}, L245 (1997).

 \bibitem{Fons1996}
 J. T. Fons, R. S. Schappe, and C. C. Lin,  Electron-impact excitation of the second positive band system ($C^3\Pi_u \to B^3\Pi_g$) and the $C^3\Pi_u$ electronic state of the nitrogen molecule, Phys. Rev. A \textbf{53}, 2239 (1996).

  \bibitem{Sprangle2011}
 P. Sprangle, J. Pe$\tilde{n}$ano, B. Hafizi, D. Gordon, and M. Scully, Remotely induced atmospheric lasing, Appl. Phys. Lett. \textbf{98}, 211102 (2011)

 \bibitem{Itikawa2006}
 Y. Itikawa, Cross Sections for Electron Collisions with Nitrogen Molecules, J. Phys. Chem. Ref. Data \textbf{35}, 31 (2006)

 \bibitem{Jones2005}
 R. J. Jones, K. D. Moll, M. J. Thorpe, and J. Ye, Phasecoherent frequency combs in the vacuum ultraviolet via high-harmonic generation inside a femtosecond enhancement cavity, Phys. Rev. Lett. \textbf{94}, 193201 (2005).
%26
  \bibitem{Gohle2005}
C. Gohle, T. Udem, M. Herrmann, J. Rauschenberger, R. Holzwarth, H. A. Schuessler, F. Krausz, and T. W. H$\ddot{a}$nsch, A frequency comb in the extreme ultraviolet, Nature \textbf{436}, 234 (2005).

 \bibitem{Lee2011}
J. Lee, D. R. Carlson, and R. J. Jones, Optimizing intracavity high harmonic generation for XUV fs frequency combs, Opt. Express \textbf{19}, 23315 (2011).

  \bibitem{Mills2012}
A. K. Mills, T. J. Hammond, M. H. C. Lam, and D. J. Jones, XUV frequency combs via femtosecond enhancement cavities, J. Phys. B \textbf{45}, 142001 (2012).

 \bibitem{Ozawa2015}
A. Ozawa, Z. G. Zhao, M. Kuwata-Gonokami, and Y. Kobayashi, High average power coherent vuv generation at 10 MHz repetition frequency by intracavity high harmonic generation, Opt. Express \textbf{23}, 15107 (2015).

 \bibitem{Zhang2020}
J. Zhang, L. Hua, Z. Chen, M. Zhu, C. Gong, and X. Liu, Extreme ultraviolet frequency comb with more than 100 $\mu$W average power below 100 nm, Chin. Phys. Lett. \textbf{37}, 124203 (2020).

  \bibitem{Yost2009}
M. A. R. Reber, Y. Chen, and T. K. Allison, Cavity-enhanced ultrafast spectroscopy: ultrafast meets ultrasensitive, Optica \textbf{3}, 311 (2016).

 \bibitem{Benko2015}
C. Benko, L. Hua, T. K. Allison, F. Labaye, and J. Ye,  Cavity-Enhanced Field-Free Molecular Alignment at a High Repetition Rate,  Phys. Rev. Lett. \textbf{114}, 153001 (2015).

  \bibitem{Hogner2017}
M. H$\ddot{o}$gner, V. Tosa, and I. Pupeza, Generation of isolated attosecond pulses with enhancement cavities¡ªa theoretical study, New J. Phys. \textbf{19}, 033040 (2017).



 %\bibitem{Liu2016}
 %X. Liu, W. Cheng, M. Petrarca, and P. Polynkin,  Measurements of fluence profiles in femtosecond laser filaments in air,  Opt. Lett. \textbf{41}, 4751 (2016).


 \bibitem{Corder2018}
C. Corder, P. Zhao, J. Bakalis, X. Li, M. D. Kershis, A. R. Muraca, M. G. White, and T. K. Allison, Ultrafast extreme ultraviolet photoemission without space charge, Struct. Dyn. \textbf{5}, 054301 (2018).

 \bibitem{Na2019}
M. X. Na, A. K. Mills, F. Boschini, M. Michiardi, B. Nosarzewski, R. P. Day, E. Razzoli, A. Sheyerman, M. Schneider, G. Levy, S. Zhdanovich, T. P. Devereaux, A. F. Kemper, D. J. Jones, and A. Damascelli, Direct determination of mode-projected electron-phonon coupling in the time domain, Science \textbf{6}, 1231 (2019).
%31
 \bibitem{Allison2011}
T. K. Allison, A. Cing$\ddot{o}$z, D. C. Yost, and J. Ye,  Extreme Nonlinear Optics in a Femtosecond Enhancement Cavity,  Phys. Rev. Lett. \textbf{107}, 183903 (2011).

 \bibitem{Porat2018}
 G. Porat, C. M. Heyl, S. B. Schoun, C. Benko, N. Dorre, K. L. Corwin, and J. Ye,  Phase-matched extreme-ultraviolet frequency-comb generation,  Nat. Photonics \textbf{12}, 387 (2018).

 \bibitem{Zhang2019}
 J. Zhang, L. Hua, S. Yu, Z. Chen, and X. Liu,  Femtosecond enhancement cavity with kilowatt average power,  Chin. Phys. B \textbf{28}, 044206 (2019).

   \bibitem{Hariri2014}
 A. Hariri, and S. Sarikhani,  Amplified spontaneous emission in N$_2$ lasers: Saturation and bandwidth study,  Opt. Commun. \textbf{318}, 152 (2014).

  \bibitem{Plenge2009}
J. Plenge, A. Wirsing, C. Raschpichler, M. Meyer, and E. Ruhl,  Chirped pulse multiphoton ionization of nitrogen: Control of selective rotational excitation in N$_2^+(B^2\Sigma_u^+)$,  J. Chem. Phys. \textbf{130}, 244313 (2009).

%36

  \bibitem{Guo1998}
  C. Guo, M. Li, J. P. Nibarger, and G. N. Gibson, Single and double ionization of diatomic molecules in strong laser fields, Phys. Rev. A \textbf{58}, R4271 (1998).

  \bibitem{Miller1988}
D. R. Miller, in Atomic and Molecular Beam Methods Vol. 1 (ed. Scoles, G.) Ch. 2 (Oxford Univ. Press, Oxford, 1988).

  \bibitem{Tong2005}
X. Tong and C. D. Lin, Empirical formula for static field ionization rates of atoms and molecules by lasers in the barrier-suppression regime, J. Phys. B \textbf{38}, 2593 (2005).

 \bibitem{Danylo2019}
R. Danylo, X. Zhang, Z. Fan, D. Zhou, Q. Lu, B. Zhou, Q. Liang, S. Zhuang, A. Houard, A. Mysyrowicz, E. Oliva, and Y. Liu,  Formation Dynamics of Excited Neutral Nitrogen Molecules inside Femtosecond Laser Filaments,  Phys. Rev. Lett. \textbf{123}, 243203 (2019).

 \bibitem{Bucksbaum1986}
 P. H. Bucksbaum, M. Bashkansky, R. R. Freeman, T. J. McIlrath, and L. F. DiMauro,  Suppression of multiphoton ionization with circularly polarized coherent light,  Phys. Rev. Lett. \textbf{56}, 2590 (1986).

 \bibitem{Corkum1989}
 P. B. Corkum, N. H. Burnett, and F. Brunel,  Above-threshold ionization in the long-wavelength limit,  Phys. Rev. Lett. \textbf{62}, 1259 (1989).
%41
 \bibitem{Zheng2020}
 W. Zheng, Z. Miao, C. Dai, Y. Wang, Y. Liu, Q. Gong, and C. Wu, Formation Mechanism of Excited Neutral Nitrogen Molecules Pumped by Intense Femtosecond Laser Pulses, J. Phys. Chem. Lett. \textbf{11}, 7702 (2020).

 \bibitem{Becker2002}
W. Becker, F. Grasbon, R. Kopold, D. B. Milo$\check{s}$evic, G. G. Paulus, and H. Walther, Above-threshold ionization: from classical features to quantum effects, Adv. At. Mol. Opt. Phys. \textbf{48}, 35 (2002).
  %\bibitem{Xu2018}
  %B. Xu, S. Jiang, J. Yao, J. Chen, Z. Liu, W. Chu, Y. Wan, F. Zhang, L. Qiao, R. Lu, Y. Cheng, and Z. Xu,  Free-space $N_2^+$ lasers generated in strong laser fields: the role of molecular vibration,  Opt. Express \textbf{26}, 13331 (2018).

% \bibitem{Lei2017}
 %M. Lei, C. Wu, Q. Liang, A. Zhang, Y. Li, Q. Cheng, S. Wang, H. Yang, Q. Gong, and H. Jiang,  The fast decay of ionized nitrogen molecules in laser filamentation investigated by a picosecond streak camera,  J. Phys. B: At. Mol. Opt. Phys. \textbf{50}, 145101 (2017).

 %\bibitem{Yao2018}
 %J. P. Yao, W. Chu, Z. X. Liu, B. Xu, J. M. Chen, and Y. Cheng, % Generation of Raman lasers from nitrogen molecular %ions driven by ultraintense laser fields,} % New J. Phys.}} 20, 9 (2018).

 %\bibitem{Bey1968}
 %P. P. Bey, J. F. Giuliani, and H. Rabin,  LINEAR AND %CIRCULAR POLARIZED LASER RADIATION IN OPTICAL THIRD HARMONIC GENERATION,}  Phys. Lett. A A}} %\textbf{26}, 128-& (1968).

 \bibitem{Xu2011APL}
  H. Xu, A. Azarm, and S. L. Chin,  Controlling fluorescence from N$_2$ inside femtosecond laser filaments in air by two-color laser pulses,  Appl. Phys. Lett. \textbf{98}, 141111 (2011).

 \bibitem{Sprengers2004}
  J. P. Sprengers, A. Johansson, A. L'Huillier, C. G. Wahlstrom, B. R. Lewis, and W. Ubachs,  Pump-probe lifetime measurements on singlet ungerade states in molecular nitrogen,  Chem. Phys. Lett. \textbf{389}, 348 (2004).

 \bibitem{Peng2019}
  P. Peng, C. Marceau, M. Herve, P. B. Corkum, A. Y. Naumov, and D. M. Villeneuve,  Symmetry of molecular Rydberg states revealed by XUV transient absorption spectroscopy,  Nat. Commun. \textbf{10}, 5269 (2019).



\end{thebibliography}
\end{document}